\documentclass[a4paper,10pt,twoside,twocolumn,final]{JICS_LaTexGuidelines}

\usepackage[mincitenames=1,maxcitenames=1,maxbibnames=99,backend=biber,isbn=false,url=false,giveninits=true]{biblatex} %

\RequirePackage{geometry} 
\usepackage{cleveref}
\usepackage{graphicx}
\usepackage{booktabs}
\usepackage{ifthen}
\newboolean{showcomments}
\setboolean{showcomments}{true}

\usepackage{comment}

\begin{document}
\title{A Survey on Domain-Specific Memory Architectures}

\author{Stephanie Soldavini and Christian Pilato\\\vspace{1.2em}\small 	Dipartimento di Elettronica Informazione e Bioingegneria, Politecnico di Milano, Italy\\ e-mail: stephanie.soldavini@polimi.it, christian.pilato@polimi.it
}

\markboth{Journal of Integrated Circuits and Systems, vol. 16, n. 2, 2021}%
{SOLDAVINI et al.: A Survey on Domain-Specific Memory Architectures}%
\renewcommand*\footnoterule{}%
\footnotetext{Digital Object Identifier 10.29292/jics.v16i2.509}%
\maketitle

\begin{abstract}
The never-ending demand for high performance and energy efficiency is pushing designers towards an increasing level of heterogeneity and specialization in modern computing systems. In such systems, creating efficient memory architectures is one of the major opportunities for optimizing modern workloads (e.g., computer vision, machine learning, graph analytics, etc.) that are extremely data-driven. However, designers demand proper design methods to tackle the increasing design complexity and address several new challenges, like the security and privacy of the data to be elaborated.
This paper overviews the current trend for the design of domain-specific memory architectures. Domain-specific architectures are tailored for the given application domain, with the introduction of hardware accelerators and custom memory modules while maintaining a certain level of flexibility. We describe the major components, the common  challenges, and the state-of-the-art design methodologies for building domain-specific memory architectures. We also discuss the most relevant research projects, providing a classification based on our main topics.
\end{abstract}

\begin{indexterms}
	Domain-Specific Architecture; Memory; Heterogeneous System.
\end{indexterms}

\section{Introduction}

Heterogeneous architectures can achieve high performance due to the introduction of specialized accelerators. The use of these components can also mitigate the \textit{dark silicon} problem which prevents the entire chip to be active at the same time.
Indeed, specialized accelerators are more efficient because they are tailored for the functionality to be performed, with a customized datapath and memory architecture, and can be also turned off when unused. Heterogeneous architectures are increasingly used in a variety of domains, from edge devices to data centers~\cite{pilato2021everest}, and for almost any modern application.
Most of these applications require the elaboration of huge data sets, often with parallelizable solutions. Current design methods for hardware accelerators focus on optimizing the computation, but memory and communication are a bottleneck, limiting the performance of the systems and, in many cases, increasing the power consumption even beyond the traditional CPU-based solutions. Data-intensive applications must efficiently coordinate data transfers, storage, and computation, even with different requirements. For example, machine learning applications often require regular data transfers to make training or inference on the model, while more irregular applications like graph analytics are more data dependent, demanding efficient methods to access the data upon request.
There is no one-fits-all solution and each designer needs to build the proper architecture based on specific requirements, domains, and constraints. Memory architectures range from classic memory hierarchies to fully customized solutions. 

First, the designer needs to determine which type of architecture better suits the application domain and the requirements. On one hand, specialized architectures achieve the best performance but have limited flexibility. They are often deployed on FPGAs because the designer can create specialized solutions and then reconfigure the device to implement a new functionality. On the other hand, general-purpose architectures are more reusable, sacrificing performance and consuming more energy. Domain-specific architectures are an interesting compromise. They gain efficiency from specialization and performance from parallelism~\cite{Dally20}. They can also be reused across multiple applications, increasing flexibility. For this reason, the creation of custom ASIC chips is more affordable for these architectures. However, the design of such architectures is complex since, in many cases, it also requires methods to specify the different functionalities to be executed and to configure the runtime execution accordingly. 

This paper presents a comprehensive survey on domain-specific memory architectures. We present the fundamental concepts for building domain-specific solutions for these memory architectures, along with the main components. We discuss the challenges that motivate the use of such architectures and how existing design methods, like high-level synthesis (HLS)~\cite{nane_tcad_2016}, can support the designers in the creation of these architectures.
We also present an overview of the most relevant projects in the area, providing a taxonomy with respect to the main topics that we discuss (i.e., components, challenges, and design methods). Finally, we discuss the open challenges that can drive future researchers to more innovation in the area of domain-specific architectures.

\section{Domain-Specific Memory Architectures}

Memory architectures are a key element in the design of every computing system. They provide the infrastructure not only to store the data to be elaborated (through physical memories) but also to coordinate the data transfers and reduce the communication latency (through specific communication modules). Indeed, depending on the nature of the target application(s) and the data to be stored, it is possible to reorganize the architectures to be more efficient (\textbf{specialized memory architectures}). In many cases, the same principles can be applied to several similar applications, increasing the flexibility and the reusability of the architecture (\textbf{domain-specific memory architectures}). While this approach is more interesting, it requires a trade-off in flexibility and performance or resources.

Common architectures for general-purpose processors are based on the concept of \textit{memory hierarchy}: since the memory components may have different latency and storage capabilities, they are organized in a way that the computing logic has direct access to fast (but small) memories, while data transfers with large (but slow) memories are transparently executed in hardware. The memory hierarchy can be organized in several \textit{levels} and, in case of multiple components, only the last levels are usually shared. While this approach allows each component to have some ``private'' data copies for fast access, it can create coherence issues that require proper protocols to guarantee correct execution~\cite{Giri18}. 

\begin{figure}
\centering
\includegraphics[width=0.7\columnwidth]{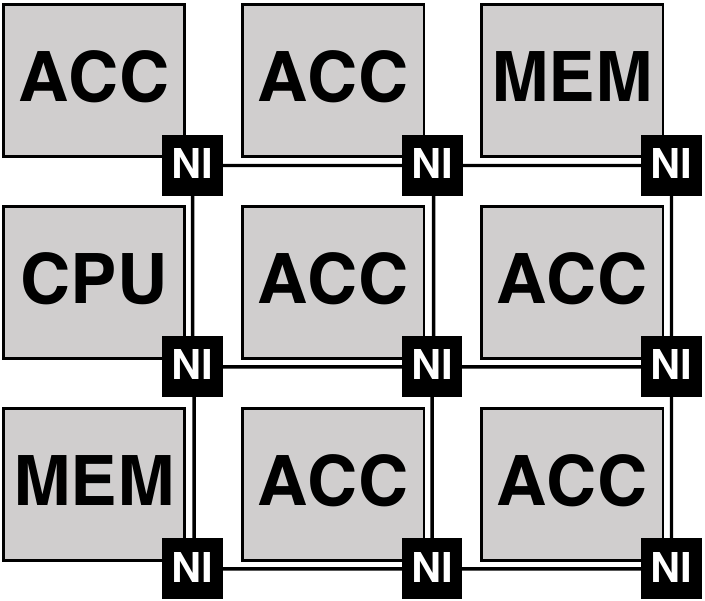}
\caption{Example of NoC-based architecture with several accelerators and memory controllers}\label{fig:architecture}
\end{figure}

With the advent of \textbf{heterogeneous architectures}, this paradigm demands extensions to manage and coordinate the accesses from different components, along with the proper design of local, on-chip memory. Besides cache coherence protocols, the designers have to manage multiple components potentially running and making requests at the same time. This creates congestion on the interconnection infrastructure, demanding more scalable systems like network-on-chip solutions~\cite{Mantovani2020}. An example of such architectures is shown in Figure~\ref{fig:architecture} and includes two memory controllers. Indeed, multiple memory controllers can help distribute the communications although they need a proper support for data allocation~\cite{Mantovani2016}. These architectures are more efficient for large, loosely-coupled accelerators that execute specific kernels of the applications. While these architectures have large benefits for specific workloads, they must be redesigned when used for other workloads.

Dataflow architectures are extremely efficient in case of large data sets with relatively simple computation, like for example machine-learning applications. In this case, data are streamed inside the architecture and \textit{buffers} may be required among the computational components. Examples of efficient dataflow architectures are the \textbf{systolic arrays}, where an array of (simple) processing elements can efficiently elaborate parallel incoming data.

\section{Overview of Memory Components}\label{sec:components}

In this section, we describe the major components that we can find in memory architecture, showing how specialization can impact their design or use.

\subsection{Off-Chip Memories}
Off-chip memory is generally used to store large amounts of data that cannot fit inside the chip. Based on the application and the use of the data, it is possible to use different technologies for storing the data.
Dynamic random-access memory (DRAM) is a type of low-cost, high-density volatile memory typically used for main memory in processing systems. DRAM cells need to be periodically refreshed, and therefore consume a relatively high amount of power. 

Volatile memory needs power to maintain data. Non-volatile memory, in contrast, can retain data while powered off. Non-volatile memory is typically used for long-term storage and common types are flash memory (such as solid-state drives and NAND flash) or magnetic memory (such as hard-disk drives and floppy disks). Non-volatile flash memory can retain data for years, but degrades at around $10^3$ to $10^5$ write cycles, while DRAM can sustain around $10^{15}$ cycles. DRAM also has a much lower access latency of $20 - 50$ns for read and write operations compared to NAND flash with a $25-125\mu$s read latency and $1-5$ms write latency \cite{Zambelli2017}. These features make volatile DRAM much more suitable for high access main memory and non-volatile flash more suitable for long term, high volume storage.

\subsection{On-Chip Memories}

On-chip memory is used to provide fast data access to the computational components. 
Static random-access memory (SRAM) is also volatile memory, but is more expensive and lower density than DRAM. SRAM, however, is faster than DRAM and is therefore well suited for on-chip memories where speed is more important than area and cost. For example, in 45nm technology with a 4GHz clock, accesses to SRAM take 8 cycles while accesses to DRAM take 24 cycles \cite{Wu2009}. Under other circumstances, the relative latency between SRAM and DRAM could be up to an order of magnitude.
SRAM has a static nature and therefore consumes less power than DRAM, since it does not need to be refreshed periodically. Due to the lower density of SRAM, its area is predominant in the components, therefore its optimization is extremely important~\cite{Pilato2017}.

SRAM physical banks can be reorganized in modules that can provide different on-demand accesses for hiding communication latency or multi-bank accesses for exploiting fine-grained data parallelism. A \textbf{cache} is common in general-purpose processors to create a memory hierarchy and to hide the latency of communication with DRAM. Data are autonomously moved across cache levels exploiting spatial and temporal locality principles. However, this introduces coherency issues in case of multiple components. The cache for an accelerator could interface with the memory hierarchy using different levels of coherency, including non-coherent, coherent with the last-level cache (LLC), or fully-coherent.
A non-coherent cache is a private memory space which uses direct memory access (DMA) to access main memory. This is simple to implement but requires many accesses to main memory, costing in performance and power.
LLC-coherent cache is the same as non-coherent except the DMA is to the last-level cache instead of main memory. This can be more power and performance efficient than non-coherent, but suffers in the case of many contending accelerators or irregular access patterns.
Fully-coherent cache implements a private cache for each accelerator with a coherence protocol such as MESI or MOESI. This is the most complex in terms of hardware implementation, but has the highest potential for good performance.
Caches provide access to a wide address space at the cost of variable latency. Latency-insensitive protocols must be used to ensure correct computation in case of cache misses. Reconfigurable caches can be used to adapt a single implementation to multiple workloads~\cite{Ranganathan2000}. A \textbf{scratchpad} memory is a local memory space managed by the host CPU. The host can transfer data to and from this memory, and the accelerator can access this memory at a fixed latency by assuming the host will have provided any necessary data ahead of time. 
A \textbf{private local memory} (PLM) stores any internal data structures needed by an accelerator at fixed latency. Typically these memories only store a portion of the working data set at a time, with the entirety stored in main memory. Since the PLM is not exposed to the CPU but completely managed by the accelerator, DMA is used to exchange data between the PLM and main memory. This memory access is managed by the accelerator itself, without the need for a host to orchestrate the transfers. Private local memories are also commonly used for temporary results produced by the accelerator.

\subsection{Communication and Coordination Components}\label{sec:comm_components}

The \textbf{memory controller} is a physical component that manages accesses to external memory. 
Typically this includes the logic to read and write, and also to refresh the DRAM cells. 
In the case of double-data rate (DDR) DRAM, the memory controller is more complex to handle transfers on both the rising and falling edges of the clock. 
Multichannel memory controllers coordinate parallel accesses to physically-separated DRAM devices on separate busses, allowing for a higher bandwidth. However, they also require proper allocation methods to decide how to partition the data across the different DRAM modules and access them correctly from the hardware~\cite{Mantovani2016}.
In-memory computing can accelerate the computation because it performs operations on the data directly inside the memory structure, dramatically reducing the cost of data transfers. This can be facilitated by modifications to the memory controller, which can activate the components in DRAM in atypical ways, control added in-memory computation cores, or perform such operations in front of classic DRAM modules. The latter case is also called \textit{near-memory computing}.

Additional components can be used to better coordinate data transfers. Based on the application domain, it is possible to extract important information on the behavior of the algorithm (e.g., regularity and locality of data transfers) that can be later used to customize the memory architecture and facilitate the data transfers. \textbf{Direct memory access} (DMA) allows memory access outside of the classic memory-hierarchy. Accelerators can directly interface with main memory, relieving the general-purpose processors from the burden of serving all accelerators at the same time.
A \textbf{prefetcher} is used to ``predict'' data accesses and transfer data from main memory to local storage before it is needed. This means that data is available at fixed latency when the accelerator requests it. To do so, it is usually combined with a DMA controller to perform the required data transfers. This is effective for regular access patterns, but has limitations with irregular patterns.
A PLM can be organized as a \textbf{reuse buffer} when portions of data are reused in successive operations. For instance, when filtering an image, an $n\times n$ region of pixels is needed for one calculation. When filtering an adjacent pixel in the image, most of the same region, $(n-1) \times n$, of surrounding pixels is needed. A reuse buffer would be coordinated such that only the new $n$ pixels need to be read from main memory and the overlapping region is not re-read unnecessarily. 
A \textbf{circular buffer} is a set of memory banks laid out sequentially such that accesses occur in a First-In-First-Out (FIFO) pattern. When access reaches the last bank in the sequence, it wraps around to the first bank to form a circular access pattern. Typically a circular buffer has one read ``pointer'' and one write ``pointer'' so that data can be written and read independently. This way, while a consumer is reading and processing data, a producer can be writing more data elsewhere in the buffer for the consumer to use later.
A \textbf{ping-pong buffer} is a pair of memory banks where, at any time, one bank is being read from by a consumer, and the other is being written to by a producer. For instance, if an accelerator needs to operate on a data structure, it can execute using data in the first bank, while a host CPU is loading the next data structure into the second bank. Once the accelerator is finished with the first data structure, it can begin execution using the second bank while the CPU loads again new data into the first bank. This technique allows data transfer latency to be ``hidden'' behind the accelerator execution but doubles the memory requirements.

To better manage heterogeneous resources, FPGA devices require abstraction layers. An FPGA \textbf{overlay} is a virtual architecture layer above the physical FPGA fabric. This virtual layer could abstract the FPGA to have a different configuration granularity, to behave like a CGRA architecture, or to expose a processor-like system to the programmer. The benefits of using this virtual layer include portability between FPGAs from competing vendors, reduced compile time, more rapid and intuitive debugging capabilities, and a clear separation between software and hardware concerns.

\section{Overview of the Main Challenges}

While specialization can bring significant benefits, it also introduces important design challenges. Based on the target application or domain, the designers need to determine which elements can be removed from a general-purpose architecture and which others can be introduced to improve the design. In the case of domain-specific memory architectures, the designers need also to carefully analyze the organization of the data and determine how they can be moved efficiently across the architecture. We analyze the following challenges: performance and delay, resources, energy, and programming effort. While the first three challenges are more focused on the implementation of the architecture, the last challenge is related to the use of the architecture.

\subsection{Performance and Delay}

Providing fast access to the data is one of the major goals of any memory architecture. First, each memory technology has different delays for the memory operations (e.g., DRAM vs. non-volatile technologies)~\cite{Wu2009}. The designers need to carefully decide the proper memory technology based on application requirements in terms of data persistence and access delay. In some cases, hybrid architectures are also possible to trade-off such requirements. 
In the case of specialized memory architectures, the computation portion of an accelerator can be significantly improved in the case of \textbf{fixed-latency memory accesses}. For example, high-level synthesis can schedule the memory operations more efficiently, extracting more operation-level parallelism~\cite{Pilato2011}. On the contrary, exploiting data-level parallelism requires \textbf{multiple concurrent memory accesses}, which necessitates proper techniques to avoid conflicts~\cite{Pilato2017} or logarithmic interconnects to reduce access delay~\cite{Rahimi2011}. Streaming interfaces and dataflow architectures can improve the throughput of such systems, especially for data-intensive applications. They create a pipelined computation, like in processor stages.
Another important problem in memory architectures is related to the optimization of the data transfers. For example, in multimedia applications, the size of on-chip memories is usually orders of magnitude smaller than the size of the entire data set~\cite{Mantovani2016}. Since all data cannot fit on-chip, it is usually necessary to exchange data between on- and off-chip memories. Such frequent data transfers can create a bottleneck in the communication infrastructure. Multiple memory controllers with \textbf{hardware translation units} can mitigate such problems~\cite{Mantovani2016}, but they require appropriate support during the design phase to allocate the data and create the communication logic accordingly. DMA engines and prefetchers can be used, instead, to make \textbf{parallel data transfers} and anticipate them in order to \textbf{overlap communication and computation}, which must be decoupled~\cite{Chen16,Ham17}. In all cases, such optimizations require extensive tool support for automatic design~\cite{Koep18}.

\subsection{Resource}

The cost of memory IPs usually dominates the use of resources for data-intensive architectures. For example, in embedded heterogeneous architectures, PLMs can occupy more than 90\% of the accelerator area~\cite{Pilato2017}. This cost is exacerbated in the case of multi-port memories. On one side, multi-port memory IPs grow quadratically with the number of ports~\cite{Tatsumi2000}. On the other side, it is possible to build multi-port memory elements on top of one- or two-port memory IPs, but this usually requires partitioning or duplicating the data~\cite{Pilato2017}, which leads to additional resource requirements. In addition, the steering logic, along with the logic for memory controllers and other communication engines, is not negligible (in some cases up to 20\% of the memory architecture). Similarly, other performance optimizations, like the various buffers described in \Cref{sec:components}, increase both memory and logic requirements. While the resource requirements generally increase the cost of chip manufacturing--larger area requires more silicon, making the chip more expensive--this can make the implementation of an architecture unfeasible for FPGA devices, for example when the architecture cannot fit into the available device.

While specialized memory architectures are tailored for specific applications, thus minimizing the use of resources, domain-specific memory architectures usually have larger requirements due to a more regular and flexible organization.
\textbf{Resource sharing} is a popular technique to reduce resource requirements. It allows designers to reuse the same physical resources over time. Since most data-intensive applications have a predictable behavior, designers can determine the \textit{liveness intervals} of each memory, i.e., the periods of time for which the memory holds meaningful values. With this information, they can share memories (i.e., implement them with the same physical banks) when the corresponding intervals are not overlapping. However, this optimization is technology-dependent as it requires knowing the size of the memories and the cost of the additional steering logic to make educated choices.

\subsection{Energy}

Energy consumption is a critical problem in embedded devices as they are often battery-powered. Since energy is the product of the time required for the computation and its power consumption, it is necessary to reduce both. Performance optimizations are important to reduce time, while resource optimizations can reduce the amount of logic and, in turn, the (static) power consumption. However, there are other important sources of power consumption that can be optimized.
While the power consumption of memory elements is a major concern~\cite{Pilato2017}, data movements are equally expensive~\cite{Mbak18}. For this reason, optimizations of such transfers, like the use of reuse buffers, are important to reduce the \textbf{number of off-chip accesses}. Similarly, the reuse of the same memory elements can significantly reduce resource requirements and, in turn, the corresponding static power of unused modules. Alternatively, unused modules can be left in the architecture and reused to \textbf{extend the LLC of the processors} to reduce the number of DRAM accesses.
These optimizations are even more relevant in the case of specific application domains. For example, in deep neural networks, it is possible to trade off internal memory size and memory access energy based on the network parameters and the given technology~\cite{Hsiao2018}.

\subsection{Programming Effort}

The design of custom architectures (either specialized or domain-specific) is complex because it requires the specification of different \textbf{functional and non-functional requirements} at high levels of abstraction. This leads to long development time that, in the case of domain-specific architectures, is only partially mitigated by the reusability of the system. In this scenario, traditional high-level languages, like the ones based on C/C++, have little expressiveness for timing and concurrency (necessary to correctly coordinate the components) while they are already ``implementing'' some decisions, like memory layout and accesses (limiting memory operations). Conversely, \textbf{domain-specific languages} are widely used to abstract hardware details, perform domain-specific transformations at the compiler level~\cite{Rink18}, and create architectures based on templates~\cite{Koep18}. However, deciding which information should be automatically inferred by the compiler and which must still be passed by the programmer is still an open issue.

\section{Design Methodologies}

After describing the main architectures, components, and challenges, in this section we describe the most common methodologies to design and optimize memory architectures with a focus on the specialization for a given application domain. We classify them based on the challenges discussed above. In all cases, design space exploration methods can be used to analyze alternative solutions and optimize a given metric~\cite{Piccolboni17}.

\subsection{Performance Optimization}

Operator dependencies often depend on memory accesses. When these memory accesses use addresses computed at runtime, it can be difficult to know if two accesses are to the same location and depend on one another. Compilers can perform \textbf{alias analysis} to determine if any accesses can be guaranteed to refer to different locations. If the compiler cannot make this guarantee, it should be conservatively assumed that the accesses could be to the same location and are dependent and the proper order must be maintained. A false dependence can stall execution, so effectively determining which accesses are not actually dependent can reduce stalls and increase performance. 

Moving data to an accelerator often accounts for a large portion of the overall execution time of an accelerator. 
Because memory accesses are often the slowest operations for an accelerator, separating memory and address calculation operations and data computation operations and allowing the memory portion to execute ahead of the computation portion can reduce the amount of time the computation is stalled waiting for a memory access. 
This data transfer latency can be partially hidden by \textbf{pipelining} the data transfers with the execution. While the accelerator is executing, new data can be transferred in and the previous results can be transferred out simultaneously. This method requires a ping-pong buffer, increasing the memory footprint, and coordination components that perform data transfers independently from the computation. Also, designers may use \textbf{FPGA prototyping} to analyze complex behaviors and balance communication and computation if needed~\cite{Mantovani2016_dac,Mantovani2020}.

Because data typically needs to be transferred between the host and the accelerator, the layout of the data in memory can have a huge impact on the time it takes to transfer all of the data. If the data is arranged such that it can be transferred in large bursts, it will be much more efficient than if the data is scattered and must be transferred in small batches. Specific OS modules can be used to manage such allocation~\cite{Mantovani2016_dac}.  

\subsection{Resource Optimization}

\begin{figure}[!tp]
\centering\includegraphics[width=0.9\columnwidth]{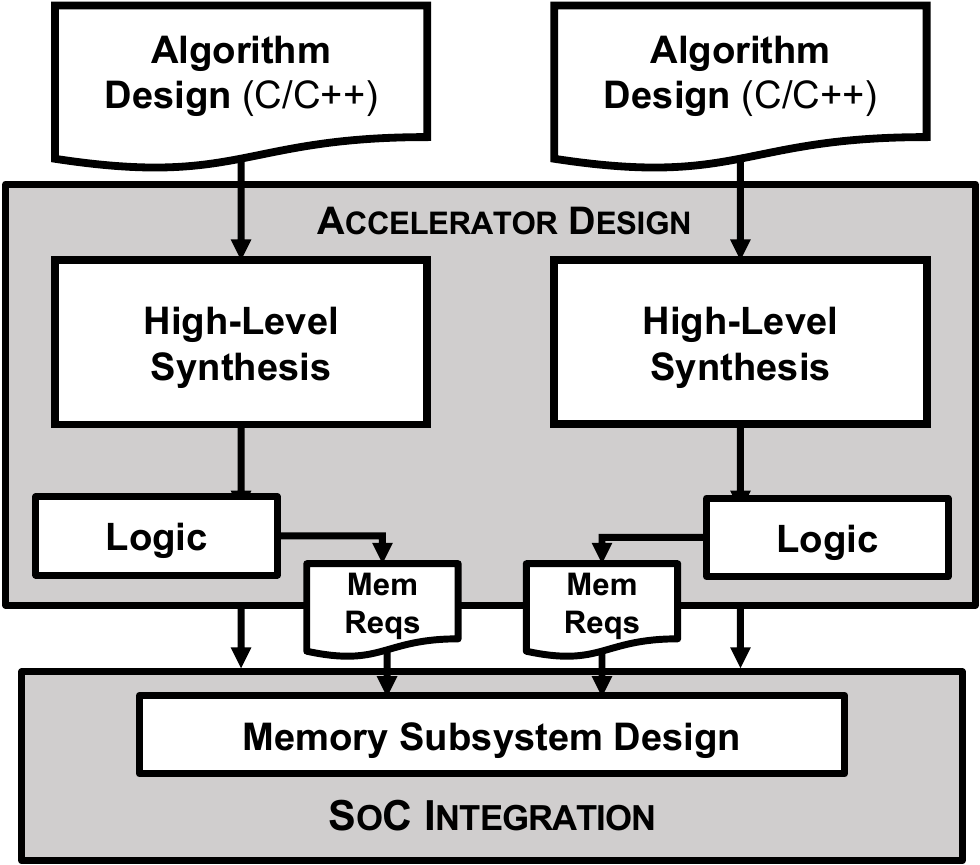}
\caption{System-level design flow for optimizing memory architectures}\label{fig:memory_design}	
\end{figure}

Because local memories contribute to a large portion of accelerator area, sharing banks when possible can help minimize the area cost. Within one accelerator, memories can be shared when local data structures have disjoint lifetimes~\cite{Pilato2017}. Across multiple accelerators, memories can be shared when the accelerators do not execute at the same time, or when the output of one accelerator is the input to the next. In this case, lightweight memory interfaces are used to wrap the physical banks and to guarantee the performance along with the correct accesses, demanding a system-level approach to identify compatibilities and exploit optimizations, like shown in Figure~\ref{fig:memory_design}. While this optimization is efficient, it limits the reusability of the banks across different applications as it constrains the accesses to the same physical banks. Accelerator memories can also be ``borrowed'' by other components such as to expand the available cache space~\cite{Cota2016}. In any of these cases, memory resource utilization can be reduced or the same amount of physical memory can be used more effectively. Similarly, regular memory systems with shared banks can provide better use of resources~\cite{Lyons12}. In this case, a dynamic layer assigns the memory banks to the accelerators.

\subsection{Energy Optimization}

Energy savings often come along with other optimizations. Reducing off-chip accesses will increase performance but will also eliminate the associated energy cost for those accesses. Reducing the number of physical memories saves resources, but also saves the static and dynamic energy required to access those memories. Fine-grained voltage regulation of the memory banks can help reduce the power consumption of the memory banks, especially when stalling for system congestion~\cite{Pilato2018}. This optimization, however, requires dual-rail memory IPs and integrated voltage regulators, making the physical implementation more complex.

\subsection{Programming Effort}

While many of these methodologies can be implemented manually, this is a painstaking process and requires intimate hardware knowledge. Automated design methods, especially based on high-level synthesis, may reduce the programming effort. High-level synthesis can automatically generate efficient memory architectures and explore alternative solutions~\cite{Pilato2011}. Also, we can use HLS to generate efficient memory architecture to make parallel accesses~\cite{Castellana2021} or optimize the representations in specific application domains~\cite{Siracusa2020}.
Because HLS typically integrates mature software compilers, many optimizations, like alias analysis, are easily done whereas by hand they are very time consuming and complex. Novel approaches can be based on a progressive lowering of multi-level compiler representations~\cite{Soldavini2021}.

\section{Taxonomy of Existing Projects}

This section aims at analyzing existing research projects that deal with the creation of specialized and domain-specific memory architectures. 

\textbf{LEAP} (Logic-based Environment for Application Programming) \cite{Adler11} is a platform for application development on reconfigurable hardware. The platform provides a scratchpad architecture which dynamically allocates and manages memory arrays. Varying levels of caching are implemented as well as the automatic generation of the communication infrastructure.
\textbf{CoRAM} (Connected RAM) \cite{Chung11} is an FPGA memory architecture designed to simplify the creation of infrastructure logic for the memory accesses of accelerators. The CoRAM architecture provides a virtualized memory space to the hardware accelerators, increasing portability and scalability and reducing design effort. 

\parskip=0pt plus 0pt

The \textbf{Accelerator Store} \cite{Lyons12} is a component designed to minimize accelerator area by managing the allocation of shared memory for many accelerators. Large, low-bandwidth, non-dependent memories are shared to save area without incurring too much performance overhead. Inter-accelerator FIFOs are also ideal candidates for sharing especially in the case of merging input and output FIFOs between accelerators. The accelerator store is shown to achieve 30\% area reduction while only incurring 2\% performance and $0-8\%$ energy overhead.
\textbf{ESP} \cite{Mantovani2020} is a research platform for heterogeneous SoC design and allows architects to rapidly prototype complex SoCs. The ESP architecture is organized as a grid of tiles, where these tiles can be processors, memory controllers, accelerators, or other auxiliary modules. The processor tiles can be any of several supported architectures with their own private L2 cache. The memory tiles each have a channel to external DRAM and a partition of the LLC. The accelerator tiles can be loosely-coupled accelerators generated using the ESP design flow or third-party accelerators can be integrated. These accelerators are able to access memory using several different DMA and cache coherence models. The auxiliary tiles are for any other shared peripheral such as the Ethernet NIC or UART. These tiles are all connected via sockets to a NoC. 
\textbf{ARTICo$^3$} \cite{Rodriguez2018} is an architecture for flexible hardware acceleration, offering dynamically-adaptable, high-performance computing and exploiting task and data-level parallelism. Accelerators access data via an optimized DMA-powered communication infrastructure. The memory structure of ARTICo$^3$ is hierarchical with three levels: global, local, and registers. Global memory is shared with the host CPU, local memory is common to all logic within an accelerator, and registers are fast access within an accelerator. ARTICo$^3$ is shown to achieve $13\times$ performance and $9.5\times$ less energy consumption compared to software-based platforms.
\textbf{PULPv2} (Parallel Ultra-Low-Power Version 2) \cite{Rossi2017} is an energy-efficient parallel SoC architecture. PULPv2 uses a heterogeneous memory architecture composed of latch-based standard cells memories (SCMs) and SRAMs. SRAMs operate at a higher voltage than the datapath, while SCMs are more comparable to the core digital logic. SCM cells, however, are much larger than SRAM cells, so a heterogeneous architecture is used to compromise. A set of lightweight MMUs is also used to manage the address space according to the workload. These MMUs dynamically divide the memory space into private and shared regions to maximize energy efficiency and performance. The SoC was shown to achieve 1 GOPS within 10 mW for a fully programmable 32-bit architecture.

The \textbf{Prefetching and Access/Execute Decoupling} framework \cite{Chen16} can be used to design accelerators which can tolerate long and variable memory latency. This is done using prefetching and access/execute decoupling. This work was shown to achieve an average of 2.28$\times$ performance speedup and 15\% energy consumption reduction in HLS-generated accelerators.
\textbf{ROCA} \cite{Cota2016} is a technique which exposes PLMs of accelerators to the LLC while the accelerator is not in use. ROCA implements the necessary overhead to enable this, including an enlarged tag array in LLC to track cache blocks stored in PLMs, logic to allow accelerators to reclaim their PLMs, logic to disable cache access to the PLM based on accelerators' activity rate, logic to coalesce memories of various sizes and expose them to the LLC as one PLM, and logic to flush dirty cache blocks in PLMs. ROCA is shown to achieve 70\% performance and 68\% energy improvements compared to regular cache of the same area. 
The \textbf{Heterogeneous Cache-Coherence} protocol \cite{Giri18} is an extension to the MESI directory-based cache coherence protocol and integrates LLC-coherent accelerators into a NoC architecture. The architecture supports non-coherent, fully-coherent, and LLC-coherent models and allows them to exist simultaneously for various accelerators and even supports runtime selection. It is shown that supporting LLC-coherent accelerators can achieve $4\times$ performance compared to non-coherent accelerators.
\textbf{DeSC} (Decoupled Supply-Compute) \cite{Ham17} is a communication management approach which separates data accesses and address calculations from value computations. The goal of this is to combine the performance and energy efficiency of scratchpad solutions with the low programmer effort and portability of cache-based solutions. DeSC also integrates a compression scheme to reduce traffic between the supply and compute devices. DeSC was shown to achieve roughly $2\times$ speedup.
\textbf{DSE} (Decoupled Storage Element) \cite{Hong2020} is a storage structure for decoupled execution in CGRAs. These DSEs are small buffers which can be chained or aligned to fit different application needs. Experimental results show that CGRA performance can be improved by an average of $2.53\times$ while saving dozens of processing elements.
The \textbf{Bayesian Cache Coherence Optimization} framework \cite{Bhardwaj2019} can determine which type of cache coherence interface to use for various accelerators in a system. This results in a performance-aware hybrid coherency interface suited for different applications. This work was shown to achieve 23\% better performance than a system where all accelerators used a single coherency model.

\textbf{Tesseract} \cite{Ahn2015} is a programmable processing-in-memory (PIM) accelerator for large-scale graph processing. Tesseract exploits the architecture of 3D-stacked memory and places a simple in-order PIM core in each partition of memory. These cores can request, over an efficient communication interface, computation on data that resides in a different memory partition. Functions are moved to data rather than the data being moved around to different cores. Tesseract was shown to achieve $13.8\times$ performance and $87\%$ energy reduction compared to a conventional state-of-the-art system.
\textbf{Gemmini} \cite{Genc2020} is a full-stack DNN accelerator generator. From a flexible hardware template, Gemmini generates a wide design-space of ASIC accelerators. Gemmini also provides a software stack and an SoC environment. 
The central unit of the Gemmini hardware template is a systolic spatial array. To maximize the rate of data moved into the scratchpad per iteration, Gemmini uses heuristics based on the loop tile sizes to determine when and how much data to move between DRAM, cache, and scratchpad memory. 
Gemmini accelerators have comparable performance to state-of-the-art, commercial DNN accelerators and up to three orders-of-magnitude speedup over high-performance CPUs. 

\textbf{COSMOS} \cite{Piccolboni17} is an automated methodology for design space exploration of accelerators. This methodology coordinates HLS and memory optimization tools together to generate a set of Pareto-optimal implementations for each component. COSMOS is shown to reduce the times the HLS tool is run by up to $14.6\times$ while exploring just as completely as an exhaustive search. 
\textbf{Mnemosyne} \cite{Pilato2017} is a tool which generates an optimized PLM architecture for accelerators. Based on the specification (sizes, number of ports) of arrays needed by an accelerator and the compatibilities of these arrays, Mnemosyne creates an architecture which shares physical memory banks while transparently exposing the required access ports to the accelerator. The arrays can either be address-space compatible when their lifetimes are not overlapping or memory-interface compatible when two reads or two writes never happen concurrently. When arrays are address-space compatible, they can share the same physical memory space. When they are memory-interface compatible, the physical ports of a memory bank can be shared even while the data occupies disjoint memory locations. Mnemosyne is shown to achieve up to 45\% memory cost savings for single accelerators and up to 55\% memory cost savings when sharing across multiple accelerators.
\textbf{Spatial} \cite{Koep18} is a domain-specific language for high level descriptions of accelerators. The abstractions target increased programmer productivity and accelerator performance. Particular features of the language enable pipeline scheduling, automatic memory banking, and automated design tuning. Spatial provides templates for various memories, including read-only lookup-tables, scratchpads, line buffers, FIFOs and LIFOs, registers, and register files. Off-chip memory access is also enabled through use of ``shared'' memories allocated by the host CPU and through several communication interfaces such as arguments or streams. Spatial has been demonstrated to achieve average speedup of $2.9\times$ compared to SDAccel with 42\% less code. 

\textbf{RELISH} (Runahead Execution of Load Instructions via Sliced Hardware) \cite{Fleming2017} is a LegUp HLS optimization pass which constructs a ``pslice'' (precomputation slice) for an accelerator. A ``pslice'' is an executable portion of an original program which only includes certain operations, in this case every long latency global load in the accelerated function. This pslice runs in parallel to the rest of the function, executing loads as early as possible and placing responses into a FIFO for the original circuit to use later. RELISH was shown to generate circuits with $1.05-1.69\times$ speedup with $1.15\times$ area overhead. \textbf{LegUp-NoC} \cite{Islam2018} is a LegUp HLS compiler pass which inserts NoCs in between datapaths and memories. This pass handles loops with indirect memory access, provides a performance and resource tuning framework, and abstracts away the need for NoC expertise during implementation. Experimental results show $5-20\times$ speedup with $20-30\%$ area overhead. 
\textbf{NACHOS} \cite{Vedula2018} is a methodology for disambiguating memory access aliasing for accelerators. At compile time, accesses are classified as NO alias (independent accesses), MUST alias (ordering must be enforced), or MAY alias (compiler uncertain). Conservatively, both MUST and MAY alias accesses should be serialized. NACHOS is a hardware comparator used to assist in disambiguating MAY aliases dynamically. NACHOS is able to achieve comparable performance to an optimized Load-Store-Queue, and achieved up to 70\% performance improvement in some benchmarks. NACHOS contributes to $\simeq6\%$ energy consumption in a system whereas an optimized LSQ contributes 27\% energy consumption.

\textbf{Data Offloading} may include dispatching methods for FPGA accelerators of streaming applications~\cite{Mbak18}. One method uses zero-copy data transfers and scratchpads, the next uses zero-copy with shared copy engines across different accelerators and local external memory, and the last uses the CPU's memory management unit to decode the physical address of user pages and uses scatter-gather transfers with scratchpads. The first method increases energy efficiency, while all methods increase scalability. 

\begin{table*}[htbp]
    \centering
    \caption{Research Project Taxonomy}\label{fig:taxonomy}
    \begin{tabular}{cll} \toprule
        & & Projects \\ \midrule
        & Cache & 
            \cite{Adler11}
            \cite{Bhardwaj2019} 
            \cite{Cota2016}
            \cite{Genc2020} 
            \cite{Giri18} 
            \cite{Mantovani2020}
            \\
        & Scratchpad & 
            \cite{Adler11}
            \cite{Genc2020} 
            \cite{Koep18}
            \cite{Mbak18} 
            \\
        Components 
        & PLM & 
            \cite{Cota2016}
            \cite{Pilato2017} 
            \\
        & DMA & 
            \cite{Giri18} 
            \cite{Mantovani2020}
            \cite{Rodriguez2018}
            \\
        & Prefetcher & 
            \cite{Chen16} 
            \cite{Fleming2017} 
            \\ \midrule
        & Performance & 
            \cite{Ahn2015}
            \cite{Bhardwaj2019}
            \cite{Chen16} 
            \cite{Cota2016}
            \cite{Fleming2017}
            \cite{Genc2020} 
            \cite{Giri18} 
            \cite{Ham17} 
            \cite{Hong2020} 
            \cite{Islam2018}
            \\
        & & 
            \cite{Koep18} 
            \cite{Peltenburg2019} 
            \cite{Rodriguez2018}
            \cite{Seshadri2017} 
            \cite{Vedula2018}
            \\
        & Resources &
            \cite{Hong2020} 
            \cite{Lyons12}
            \cite{Pilato2017}
            \\
        & Energy & 
            \cite{Ahn2015} 
            \cite{Chen16}
            \cite{Cota2016}
            \cite{Ham17} 
            \cite{Mbak18} 
            \cite{Rodriguez2018}
            \cite{Rossi2017} 
            \cite{Seshadri2017}
            \cite{Vedula2018}
            \\
        Challenges 
        & Programmer Effort (Dev Time/Complexity) & 
            \cite{Adler11}
            \cite{Chung11}
            \cite{Ham17} 
            \cite{Koep18}
            \cite{Mantovani2020}
            \cite{Mbak18} 
            \cite{Piccolboni17} 
            \cite{Rodriguez2018}
            \\ 
        & Programmer Effort (Reorg Algorithm) & 
            \cite{Dally20} 
            \cite{Hsiao2018} 
            \cite{Cota2016}
            \\ 
        & Programmer Effort (Automation) & 
            \cite{Chen16} 
            \cite{Fleming2017} 
            \cite{Genc2020} 
            \cite{Islam2018} 
            \cite{Koep18} 
            \cite{Mantovani2020}
            \cite{Piccolboni17} 
            \\ 
        & Programmer Effort (DSE) &
            \cite{Bhardwaj2019}
            \cite{Genc2020} 
            \cite{Koep18}
            \cite{Piccolboni17} 
            \\ \midrule
        & Alias Analysis & 
            \cite{Vedula2018} 
            \\
        & Data Transfer Pipelining & 
            \cite{Fleming2017}
            \\
        Design  
        & Memory Sharing and Borrowing & 
            \cite{Cota2016} 
            \cite{Lyons12}
            \cite{Pilato2017}
            \\
        Methodologies 
        & Decoupling Memory/Compute & 
            \cite{Chen16} 
            \cite{Fleming2017}
            \cite{Ham17} 
            \cite{Hong2020} 
            \cite{Piccolboni17} 
            \\
        & Compiler Support & 
            \cite{Fleming2017}
            \cite{Islam2018}
            \cite{Vedula2018}
			\\
        & Data Layout in Memory & 
            \cite{Peltenburg2019} 
            \\
        \bottomrule\\
    \end{tabular}
\end{table*}

\textbf{Fletcher} \cite{Peltenburg2019} is an FPGA acceleration framework which uses the Apache Arrow in-memory format. The Arrow project defines a columnar in-memory format optimized for big data applications and offers communication libraries for at least eleven common programming languages. By utilizing this format, Fletcher is able to use any of these supported languages for FPGA acceleration. Fletcher is shown to accelerate applications $1.3 - 49\times$ compared to serialized hardware accelerated solutions.
\textbf{Ambit} \cite{Seshadri2017} is an Accelerator-in-Memory for bulk bitwise operations. With only minimal changes to the sense amplifiers present in DRAM, bitwise operations can be done on entire rows of DRAM. Ambit does not change the DRAM interface and therefore can be directly connected to the system memory bus. Ambit was shown to achieve $32\times$ throughput and $35\times$ energy improvement of bulk bitwise operations compared to a state-of-the-art system.

\section{Final Discussion}

Table~\ref{fig:taxonomy} shows a classification of the most relevant research projects based on the topics we discussed in the previous sections. Such taxonomy shows that performance optimization is predominant, especially with the use of scratchpads and caches, along with DMA engines to hide communication latency. 
In general, optimizing the performance has a corresponding effect on energy. On the other hand, technology improvements can bring significant benefits~\cite{Rossi2017} but sometimes have limited applicability due to technology constraints. Area savings are also gaining attention in the recent years especially for FPGA implementations to accommodate accelerators in devices with limited resources~\cite{Lyons12,Pilato2017,friebel2021domainspecific}.

While several projects primarily target programmer effort with abstractions to the programming model and despite the large interest in such architectures, automatic design methods are mostly immature, leaving most of the effort to the hardware designers. There is still a huge gap between software and hardware designers. We strongly believe this analysis can motivate researchers to propose novel methods to combine the components for the given application domain.

\section*{Acknowledgments}

This work has been partially funded by the Horizon 2020 EU Research \& Innovation Programme under grant agreement No. 957269 (EVEREST project).

\renewcommand{\bibfont}{\footnotesize}
\printbibliography

\end{document}